\documentclass[aps,pre,reprint]{revtex4-2}
\usepackage{graphicx}
\usepackage[utf8]{inputenc}
\usepackage{dcolumn}
\usepackage{bm}
\usepackage[utf8]{inputenc}
\usepackage[colorlinks=true, allcolors=blue]{hyperref}
\usepackage{listings}
\usepackage{graphicx}
\usepackage{xcolor}

\begin{document}

\title{The quest for new materials: the network theory and machine learning perspectives}
\author{Jacopo Moi}
\thanks{These authors contribute equally to the work}
\affiliation{%
DSMN, Ca'Foscari University of Venice, Italy
}
\affiliation{%
RARA Foundation - Sustainable Materials and Technologies, Venice, Italy}
\author{Davide Spallarossa}
\thanks{These authors contribute equally to the work}
\affiliation{%
DSMN, Ca'Foscari University of Venice, Italy
}
\affiliation{%
RARA Foundation - Sustainable Materials and Technologies, Venice, Italy}
\author{Stefano Bonetti}
\thanks{These authors contributed equally to this work}
\affiliation{%
DSMN, Ca'Foscari University of Venice, Italy
}
\affiliation{%
RARA Foundation - Sustainable Materials and Technologies, Venice, Italy}

\author{Raffaella Burioni}
\thanks{These authors contributed equally to this work}
\affiliation{%
University of Parma, Parma Italy
}
\affiliation{%
INFN, Sezione di Milano Bicocca, Gruppo Collegato di Parma,  Parma, Italy}
\affiliation{%
RARA Foundation - Sustainable Materials and Technologies, Venice, Italy}
 
\author{Guido Caldarelli}%
\thanks{These authors contributed equally to this work}
\affiliation{%
CNR, Institute of Complex Systems (ISC), Rome Italy
}
\affiliation{%
DSMN and ECLT, Ca'Foscari University of Venice, Italy
}
\affiliation{%
RARA Foundation - Sustainable Materials and Technologies, Venice, Italy}
\affiliation{%
London Institute for Mathematical Sciences, Royal Institution, London, UK
}

\begin{abstract}
Understanding and predicting the emergence of novel materials is a fundamental challenge in condensed matter physics,  materials science and technology. With the rapid growth of materials databases in both size and reliability, the challenge shifts from data collection to efficient exploration of this vast and complex space. A key strategy lies in a smart use of descriptors at multiple scales, ranging from atomic arrangements to macroscopic properties, to represent materials in high-dimensional abstract spaces. Network theory provides a powerful framework to structure and analyze these relationships, capturing hidden patterns and guiding discovery. Machine Learning complements this approach by enabling predictive modeling, dimensionality reduction, and the identification of promising material candidates. By integrating network-based methods with Machine Learning techniques, researchers can construct, analyze, and efficiently navigate the material space, uncovering novel materials with tailored properties. This review explores the synergy between network theory and ML, highlighting their role in accelerating materials discovery through a systematic and interpretable approach.

\end{abstract}

\maketitle

\date{\today}


\section*{Introduction}
The discovery of new materials with specific properties is a milestone in the progress of science and technology, driving innovations in the fields of energy and electronics since the age of civilization. There is little doubt that further progress in society will come from the use of new materials that will enable a more sustainable use of the planet's resources. For this reason, in $2011$ the US Administration launched the Materials Genome Initiative "to help businesses discover, develop, and deploy new materials twice as fast"\cite{whitehouse2011MGI}, to the use of Google's Tool GNoME that presented a database of 2.2 millions new crystals\cite{merchant2023scaling}, even if the novelty of those is somewhat disputed\cite{leeman2024challenges}.

It is important to highlight some aspects of the problem. Firstly, the number of possible compounds of the about one hundred elements of the periodic table is enormous \cite{kirkpatrick2004chemical}. Some authors restrict the quest by considering just small organic molecules comprising carbon, hydrogen, oxygen, nitrogen and sulfur; this smaller subset is estimated to produce a number of about $10^{60}$ different compounds  \cite{bohacek1996art}. Whatever the subset considered, the number increases when considering not just all the possible molecular arrangements with a defined geometry, but also the several possible crystalline structure belonging to same compounds. According to the various approach and simplifications some estimates indicate up to  $10^{200}$\cite{restrepo2022chemical}. Since we also ignore the structure of the configuration space to which these compounds are associated  it is clear that the exploration of such huge space is a critical question.

In this paper, we present the different approaches tested so far in exploring specific and limited subsets of the whole material space. In particular, we focus on those approaches that relevantly used machine learning and network theory. In the following, we present a  review of the research activities that utilize these two frameworks to search for new materials. The goal is to introduce these concepts to both theoreticians and experimentalists working in condensed matter physics and materials science traditionally not trained with them.  Specifically, we want to present how materials and their properties can be characterized as vectors belonging to an abstract mathematical space. Such an approach is the most convenient basis upon which the maps of the material space can be built and explored. A list of the principal databases used to build such maps, is also presented. 

The first computational approaches to material discovery (back to the 1960s) focused on using quantum mechanical methods, such as density functional theory (DFT)~\cite{sholl2022density}, to predict the properties of candidate materials with high numerical  precision. One of these first codes based on {\em ab initio} computations was the program Gaussian $70$~\cite{hohenberg1964inhomogeneous}. 

These numerical methods allowed researchers to calculate electronic structures, stability, and reactivity, offering an invaluable insight into previously unexplored materials. Early successes in computational chemistry demonstrated their potential to accelerate discovery, for instance, in the identification of new catalysts, superconductors, and materials for renewable energy applications.

The second-generation approaches introduced the use of global optimization methods, such as evolutionary algorithms, to predict possible structures. 
Indeed, the previous numerical methods were and still are computationally intensive. This limits their applicability to small datasets or to individual systems of interest. As material datasets have grown in size and complexity, those global and sometime data-driven approaches can tackle these challenges with a slower scaling of computational resources. Here, an initial input of chemical composition is transformed into predictions of the structure or set of structures that the elements are likely to form. 

The emerging third-generation approaches incorporate machine learning techniques, which, when provided with adequate data and a well-trained model, can predict composition, structure, and properties simultaneously\cite{butler2018machine}.
In this latter type of approach, both the recent development of network theory\cite{caldarelli2007scale,aziz2021towards} and the introduction of artificial intelligence (AI) have added new instruments to the search for new materials. Indeed, when dealing with these huge number of combinations, it is appropriate to consider the framework of statistical physics upon which network theory is built\cite{mandl1991statistical}.  In the case of AI, a useful tool may be represented by complex networks\cite{cimini2019statistical}, allowing quantitative mapping of a set of relations and machine learning\cite{hopfield1982neural,mitchell1997machine} that allows the detection of patterns of regularity in such a huge space.

It is useful, at this point,  to define the two key frameworks that we will discuss in detail throughout this review.
\begin{itemize}
    \item {\em Network Theory} (NT) provides a framework for understanding the complex interconnections between materials, their properties, and their underlying structures. By representing material data sets as networks, where nodes correspond to materials, and edges capture relationships such as compositional similarity or shared properties, hidden patterns can be found, similar materials clustered, and promising candidates can be identified for further investigation. This graph-based approach allows for the integration of diverse datasets, facilitating the study of large and heterogeneous materials spaces. In addition to their role as a legitimate and powerful method of analysis through mapping of materials space, networks can also be used to visualise the mapping obtained through machine learning.

    \item{\em Machine Learning} (ML) complements network theory by enabling the prediction of material properties, the identification of novel material candidates, and the optimization of synthesis routes. Through supervised and unsupervised learning algorithms, ML models can extract non-linear relationships between features, learn from experimental or computational data, and generalize to predict the behavior of previously unexplored materials. By combining these capabilities with the structural insights provided by network theory, researchers can guide experimental efforts more effectively, significantly reducing the time and resources required for materials discovery.
\end{itemize}


To explore materials space, we need, at least, to know its dimensions and to identify the key physical aspects that might affect its structure. A starting point can be to consider the data describing as many materials as possible, to gather all the relevant information.
Pursuing this long-term goal remains formidable, and it is unlikely that we will achieve convergence within the next several decades. Nevertheless, this seemingly unattainable endeavor has not deterred researchers from exploring the realm of materials. In this context, we present a selection of the studies undertaken, focusing on the fundamental frameworks that contribute to the representation of materials space.


\section*{Materials representation}
\subsection*{Methodology}
Before going into the details of computational methods, it is essential to describe the steps by which a material, as a physical object, is mapped into an abstract mathematical space. This process involves the introduction of the concept of \textit{Descriptors} and \textit{Featurization}.

\subsubsection*{Descriptors}
A material, like any other object, retains observable and measurable properties that identify it in the physical world. The first step is then to obtain a quantification of these properties in a format that retains the quantities of interest and that maximizes the information content. This is the \textit{descriptor} definition. As an example, let us take a material in which the type of atoms and their number are known. The simplest descriptor could be the brute chemical formula (eg. ZnO$_2$), which clearly qualify the material as belonging to the \textit{class} of metal oxides. Such descriptor can be further refined. If we were able to guess the position of the atoms with respect to a Cartesian system, we could now describe the material as a set of (x,y,z) tuples with respect to the Cartesian centre, also increasing the uniqueness of the representation. Furthermore, we could consider of the connectivity of atoms, we could describe the material as a graph, where the nodes are labeled according to their corresponding chemical elements and the edges between nodes represent the atom connectivity. Last but not least, we can encode the connectivity in a text string using a specific encoding such as SMILES, where ZnO$_2$ now is encoded as O=[Zn]=O. As we shall see, descriptors can be classified into ``physical'', which are defined based on first principles and correspond to fundamental, traditionally used properties and ``neural network''-based, where they are the results of a machine learning task.

\subsubsection*{Featurization and fingerprints}
The task of transforming a material into an appropriate descriptor naturally serves the purpose of distinguishing different materials from each other, according to the goal of the analysis. As introduced, descriptors can have very different encodings, which can also be of the non-linear type (e.g. chemical graph). Now, we turn to the step that involve the encoding of a descriptor to a compact numerical representation that is, transforming descriptors into numerical vectors. To remain general, this process serves two different and  strongly intertwined purposes:
(i) to represent materials as points into a multi-dimensional linear space called ``feature space''. This can achieved through a complex and physical-based processing (for example SOAP\cite{de2016comparing});
(ii) to provide in such a way a numerical input for ML. A simplest example is the conversion of a chemical formula into a specific encoded vector ({\em i.e} One-Hot vectors, where all but one bit are zero).

In both cases, this process can be modelled through  an operator $\Phi$ that maps the information (more generally an input given by an element of a given set $\chi$) into an abstract space of ``large'' dimensionality:
\begin{equation}
\Phi : \chi \rightarrow \mathbf{R}^N
\end{equation}
 Feature vectors can be built upon a range of descriptors, representing one or some properties  of the material. 
When such featurization uniquely maps the structural properties of a material to the feature space, we call it a ``\textit{fingerprint} of the material''. For example, the density of states can be embedded on a feature vector and used as a fingerprint of a specific crystal. Other examples involve the Coulomb matrix\cite{rupp2012fast} or other crystal structures\cite{schutt2014represent}.

The importance of this process lies in the properties of the feature space. Indeed, whenever it is possible to define a metric between points satisfying the prerequisites, we can define a measure of the ``distance'' between different  materials. Specifically, when an inner product can be naturally defined (see Kernel Based Method), there will be a natural notion of similarity between materials. 
Once defined a similarity metrics, we can then move in the space and for example look for duplicates, clusters, and outliers. This activity is mainly done via regression, ML and Network Theory. 

\subsection*{Physical descriptors}
The vast majority of papers selects a set of physico-chemical descriptors, which can be easily translated into their feature space.
Those descriptors are of rather different nature, ranging from emerging properties such as conductivity and magnetization \cite{singh2023physics} to structural \cite{huang2006kernel} and compositional \cite{tsekenis2024network}. Such distinction is just schematic, since this choice is not mutually excluding and several overlaps exist.  
 Whatever the descriptor might be, it must respect physical coherence with respect to invariance symmetries {\em{e.g.}} translational and rotational invariance \cite{musil2021physics}. Physico-chemical based descriptors are naturally invariant (the bandgap it is not affected by a 90 degree rotation) while structural ones have to be carefully understood in order to assess the invariance of the representation.

\subsubsection*{Macroscopic/Emergent descriptor}
As introduced, macroscopic descriptors are represented by emergent or aggregate properties arising from collective interactions within the material. Standard examples include thermal conductivity and heat capacity, bandgap and electronic conductivity, elasticity modules and fracture toughness, phase transition temperatures, dielectic constants and magnetic susceptibility.
For example, given a dataset that includes measurements for crystalline density, bandgap and thermal conductivity, a material could be featurized in this way. 
Once combined all of these into a vector, the three properties can be mapped into $\mathbf{R}^3$, where each axis describes a feature and a point describes a material indexed by the three different values of these properties. 
From a purely mathematical point of view, please note that the basis of the space is not necessarily orthogonal because there could be hidden, high-order dependencies between these features. 
Macroscopic descriptors are often experimentally measurable quantities which effectively capture collective behaviors and interactions that are critical for real-world material performance. Moreover, leveraging aggregate properties, they often simplify modeling and reduce computational demands, providing  insights into how a material will perform in practical applications. On the other hand, macroscopic descriptors are often non universal and may be tailored to particular material classes or conditions, limiting their use. 

\subsubsection*{Structural descriptors}
The simplest way to describe the spatial structure of molecules is to assign the position coordinates of each atom \cite{pronobis2020kernel}. Unfortunately, a raw encoding of this kind neglects invariance with respect to basic symmetry operations \cite{pronobis2020kernel}.
We can pass to a symmetry-conserving method by relying on chemical fingerprints which are invariant by nature. For example, Coulomb Matrix, encodes pairwise Coulombian interactions between atoms, it is not affected by rotation of a crystal-defined unit cell. 
In any case, a complete description and featurization of a given material is an open problem, and so even chemical fingerprints neglect the fundamental topological structure of the molecule \cite{brown2002topology}.
Featurizing functions are often implemented by Python packages that contain/collect featurizers at different scale levels, facilitating the workflow being compatible with most databases API's and statistic Pandas package. To give some examples as most used ones, we cite: matminer \cite{ward2018matminer} and DScribe \cite{laakso2023updates}. 
The matminer library objective is to help in data mining the properties of materials. For this reason it contains routines for creating and accessing various materials databases and to transform and featurize complex materials attributes into numerical descriptors.  
DScribe contains codes that allow transforming atomic structures into fixed-size numerical fingerprints; its descriptors are more physically coherent and complex. 

\subsection*{Neural network descriptors}
Another approach for the definition of descriptors involves leveraging Deep Neural Networks to discover new sets of features that effectively describe materials. 
Also neural network descriptors must be designed to preserve essential physical symmetries, such as translational, rotational, and permutational invariance, ensuring that models learn generalizable patterns rather than overfitting to specific representations of atomic configurations.  

There are several categories of descriptors used in neural network-based materials modeling. Some overlap with the previously presented physics-based descriptors, such as Coulomb matrices, Smooth Overlap of Atomic Positions (SOAP), and radial/angular symmetry functions that explicitly encode atomic interactions and are particularly useful in interatomic potential models. Graph-based descriptors represent materials as nodes and edges, capturing topological and connectivity information, making them well-suited for predicting complex phenomena such as electronic transport or defect formation. Learned representations, derived from deep learning architectures like convolutional neural networks (CNNs) and graph neural networks (GNNs), extract hierarchical features directly from raw structural data, often outperforming traditional handcrafted descriptors in complex property prediction tasks.  

The effectiveness of a descriptor depends on its expressiveness and computational efficiency. While highly detailed descriptors may improve predictive accuracy, they can also increase the risk of overfitting and computational overhead, particularly in large-scale materials screening. Conversely, overly simplified descriptors may fail to capture critical interactions, leading to suboptimal model performance. Hybrid approaches that combine multiple descriptor types, such as fusing local atomic environment features with global structural fingerprints, can enhance model robustness and transferability.  

The choice of descriptor has a significant impact on downstream applications, including high-throughput materials screening, inverse design, and generative modeling of novel materials. Advances in machine learning, such as self-supervised learning and equivariant neural networks, continue to refine descriptor representations, pushing the boundaries of accuracy and efficiency in materials informatics. As the field progresses, the development of domain-adaptive, interpretable, and scalable descriptors remains a key challenge in accelerating the discovery of functional materials.

An initial contribution in this area is the work of \cite{duvenaud2015convolutional}, where a Convolutional Neural Network is employed to learn molecules' representation (fingerprints) by systematically deriving higher-order structures from the raw (molecular) graph representation. Subsequent improvements followed with the works in Refs. \cite{kearnes2016molecular} \cite{gilmer2017neural}.\\

Within the framework of the ``mapping quest'', notable examples of works that leveraged neural network features obtained with different NN architectures include \cite{suzuki2022self} with a Crystal Graph Convolutional NN (CGCNN) which embeds the crystalline structure  and \cite{neporozhnii2024navigating} with a Graph Convolutional Network (GCN) ProDosNet which uses projected density of states (PDOS) data.

\subsection*{Kernel-based method}
As mentioned above, featurization of a descriptor could be hindered by the nonlinearity of its nature (e.g., chemical graph), limiting the activity to a few human-aware features. Moreover, an arbitrary mapping in a feature space does not guarantee the definition of the existence of an inner product. There may also need to move to an infinite-dimensional space to completely unravel the information stored in a particular descriptor. The concept of \textit{kernel} between data, represents an improvement in this methodology.
A kernel is a function that computes a measure of similarity between pairs of data points, often mapping the data to a higher dimensional space to make it easier to find patterns, such as for classification or clustering \cite{shawe2004kernel}. 
The transformation of features defined by $\Phi$ in addition with the inner-product define the so-called kernel $K$. This function allows us to directly compute the inner-product given an implicit description of the feature space.
A careful choice of kernels also allows us to choose the one that encodes a real valued similarity measure
between two chemical compounds\cite{pronobis2020kernel}.
\begin{equation}
    K(x_1,x_2)=\langle \Phi(x_1)\Phi(x_2) \rangle
\end{equation}
To give an example, consider the case of molecules represented as molecular graphs. The similarity between two molecules can be put in relation to graph similarity, which conserves desirable properties. The first step is to use the Simplified Molecular Input Line Entry System (SMILES), which is a specification in the form of a line notation to describe the connectivity of chemical species using short ASCII strings\cite{weininger1988smiles,weininger1989smiles} starting from the full 3D chemical structure.
From the SMILES encoding is now possible to obtain the molecular graph of the given compounds. In particular, if we decide to consider a feature map (defined by $\Phi(x)$) equal to a vector containing the number of shortest-paths of different lengths we obtain one of the possible invariant kernels known as the shortest-path kernel\cite{bogwardt2020graph}. 
The kernel inner product $\langle \Phi(x)\Phi(y) \rangle$ computes the overlap from the two inputs $x, y$, which is considered as chemical similarity between the two molecular graphs. 

\subsection*{Recent developments}
The search for new descriptors or combinations of descriptors on a macroscopic level that can be correlated to specific properties is still a very open point, as the following examples testify. 
Research on these topics is growing. For example, in \cite{purcell2023accelerating} the authors create an AI-guided workflow to find thermal insulators. They modeled the thermal conductivity of a material starting from its structural, harmonic, and the anharmonic properties. After that, they applied a feature-importance metrics to identify $16$ predicted ultra-thermal insulators. Not surprisingly, the integration of these macroscopic descriptors into machine learning models has been a significant focus in recent research. In \cite{pereti2023individual} the authors developed an approach that combined supervised classification and regression techniques to predict superconductive materials, by using macroscopic properties to improve prediction accuracy. In particular, they identified a significant shift to the ensuing critical temperature as stemming from the considered element. 
Finally, a particular interesting recent direction is that of entropic descriptors. In \cite{divilov2024disordered} the authors develop a new approach, based on the energy distribution spectrum of randomized calculations. They used it to describe the accessibility of states with equal sampling near the ground state and quantify configurational disorder stabilizing high-entropy homogeneous phases. In \cite{yang2024quantitative}, correlations between interfaces properties,  combination of order and disorder, hierarchical organization and self-assembly allow us to develop biomimetic materials.

\section*{Materials maps} 
After defining the procedure to create a feature space for materials is crucial to obtain a way to  explore it. \textit{Materials maps} allow us to do so by displaying relative distances between the materials of a selected portion of the chemical space.  

\begin{figure}[b]
    \centering
    \includegraphics[width=1\linewidth]{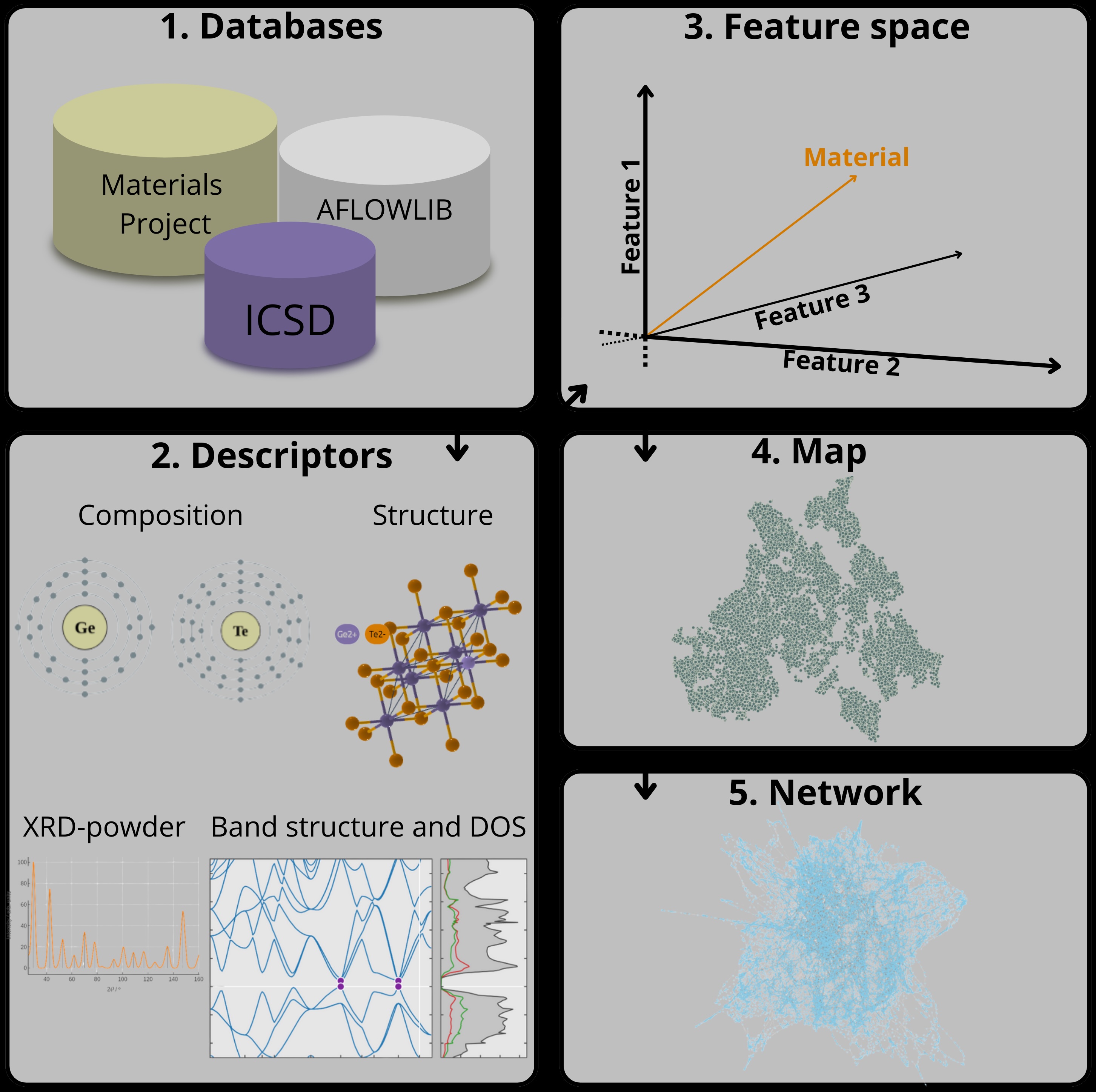}
    \label{fig:1}
    \caption{Graphical representation of the above general scheme. Numerated steps correspond to scheme sections.}
    \label{fig:schema}
\end{figure}

The common base schema to create material maps is the following (see Fig.\ref{fig:schema}):
\begin{enumerate}
    \item Select a materials database, from  specific range options (comprising a few hundred materials such as exclusively carbon structures) to more comprehensive ones (such as the Materials Project\cite{jain2013commentary}, AFLOWLIB~\cite{curtarolo2012aflowlib}, ICSD~\cite{zagorac2019recent}, etc.)
    \item Identify one or several descriptors (ranging from simple chemical composition to more complex and precise representations)
    \item Convert physical descriptors into numerical features (spanning from Boolean indicators to high-dimensional vectors)
    \item Integrate machine learning techniques, implementing clustering algorithms (DBSCAN) and dimensionality reduction methods for visualization (PCA~\cite{makiewicz1993principal}, UMAP~\cite{mcInnes2018umap}, t-SNE and sketch-map~\cite{ceriotti2013demonstrating})
    \item Establish a similarity/distance metric (general purpose one or retailed on the descriptor encoding). Progress to a material network by applying a threshold to the complete similarity matrix.
\end{enumerate}

In the following we shall show a series of selected works that illustrate the various approaches introduced in the construction of materials maps in the \textit{machine learning} classical fashion. After that we shall present the subsequent steps towards the formulation of \textit{materials networks}.

\subsection*{Machine Learning}
Hargreaves et al. \cite{hargreaves2020earth} follow carefully the above mentioned procedure to obtain a material map representation. By using only the chemical composition as descriptor, they analysed $12,623$ binary materials first and then the whole $125,627$ materials from the ICSD\cite{zagorac2019recent} ($2021$) database. Since chemical composition is a relatively simple descriptor, the importance of the encoding and the similarity metric are central to this work. The compositional vector is defined by \textit{``taking the ratio of each element in a compound assigned to the index of its respective modiﬁed Pettifor number''}\cite{hargreaves2020earth}. The earth mover's distance (EMD) is proposed in contrast to Euclidean distance. UMAP\cite{mcInnes2018umap} and PCA\cite{makiewicz1993principal} is compared, looking at clustering visualisation and the coherence of the original space distance. By using the Python package ``scikit-learn'', the code DBSCAN is performed to find cluster distributions, and it is observed that handling the entire larger database is more challenging compared to smaller binary composition datasets. 
Another interesting result is obtained in succcessive work by some of the same authors\cite{hargreaves2023database}. While studying solid-state lithium electrolytes structures, these materials have been highlighted in the previously created PCA map of the ICSD database showing clustering in this compositional map, \textit{'reflecting the connection between composition and structure'}.

\begin{figure}[t]
    \centering
    \includegraphics[width=1\linewidth]{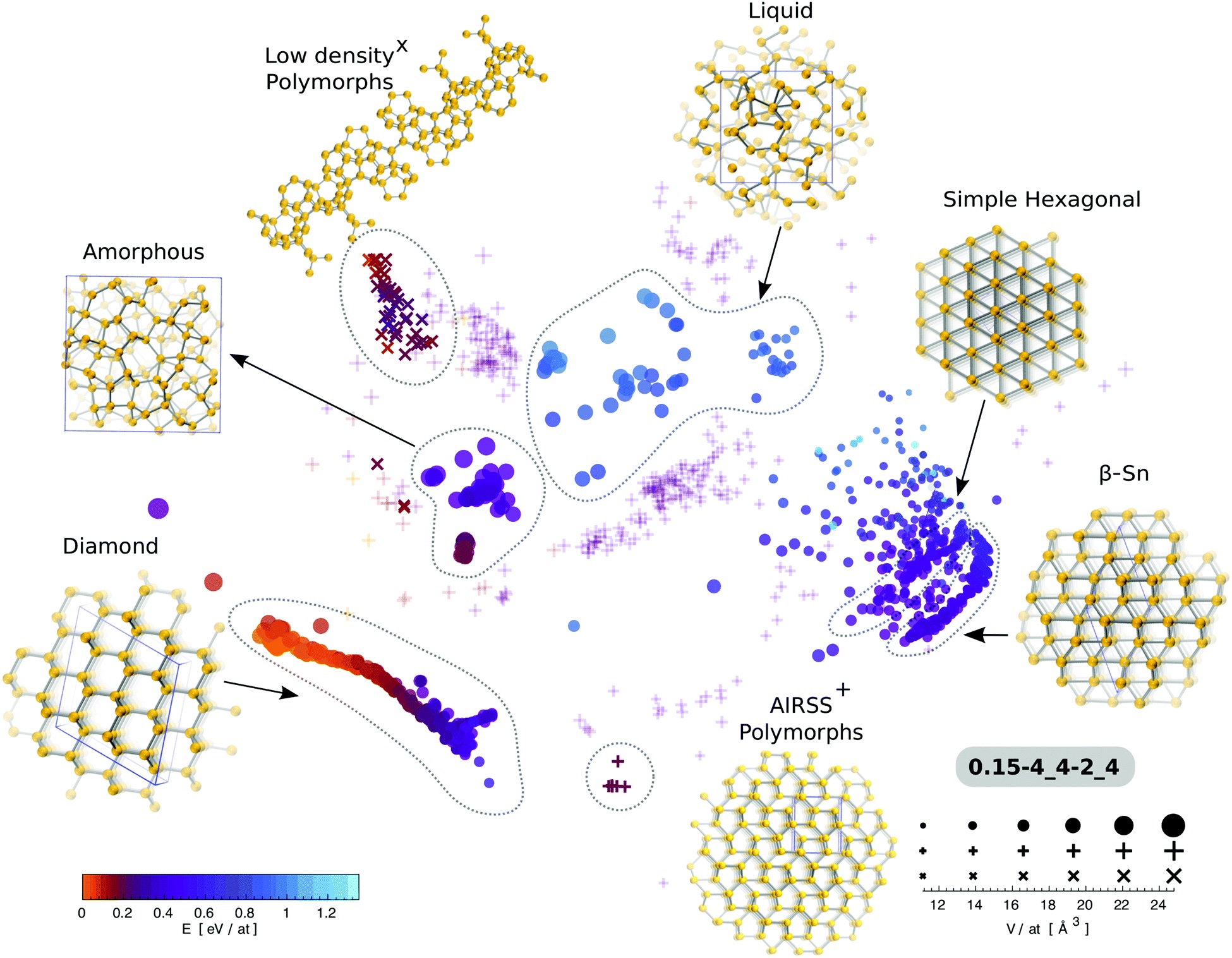}
    \caption{Figure taken from Ref\cite{de2016comparing}: "Sketch-map of $1274$ crystalline and amorphous silicon structures. The color and size of the points vary according to their atomic energy and atomic volumes, respectively. Regions of the plot which represents different phases have been outlined with dotted contours".
    \textit{Image reproduced with permission under License No.$1575322-1$. Copyright [2025], Royal Society of Chemistry}}
    \label{fig:de2016comparing}
\end{figure}

One of the state-of-the-art frameworks to describe both structural and compositional aspects is the Smooth Overlap of Atomic Positions (SOAP), \textit{'translation, rotation and permutation-invariant descriptors of groups of atoms'}\cite{de2016comparing}. Similarity is computed via a Regularized Entropy Match (REMatch) approach, which results in a great mapping of different specific materials datasets (80 configurations of $C_{60}$ [in Fig.\ref{fig:de2016comparing}], 1274 bulk silicon structures and 7211 small organic compounds from QM7b database); visualized with the sketch-map dimensionality reduction~\cite{ceriotti2013demonstrating}. Resultant maps capture structures and compositional clustering. The efficacy of the REMatch-SOAP kernels is further demonstrated by the performance of the kernel-ridge regression, where an absolute error of less than $1$ kcal/mol was achieved in predicting atomization energies through the training of $5000$ small organic molecule structures. These advanced tools have exhibited the requisite sensitivity and adaptability needed for the successful comparison of various materials classes.

\begin{figure}[t]
    \centering
    \includegraphics[width=1\linewidth]{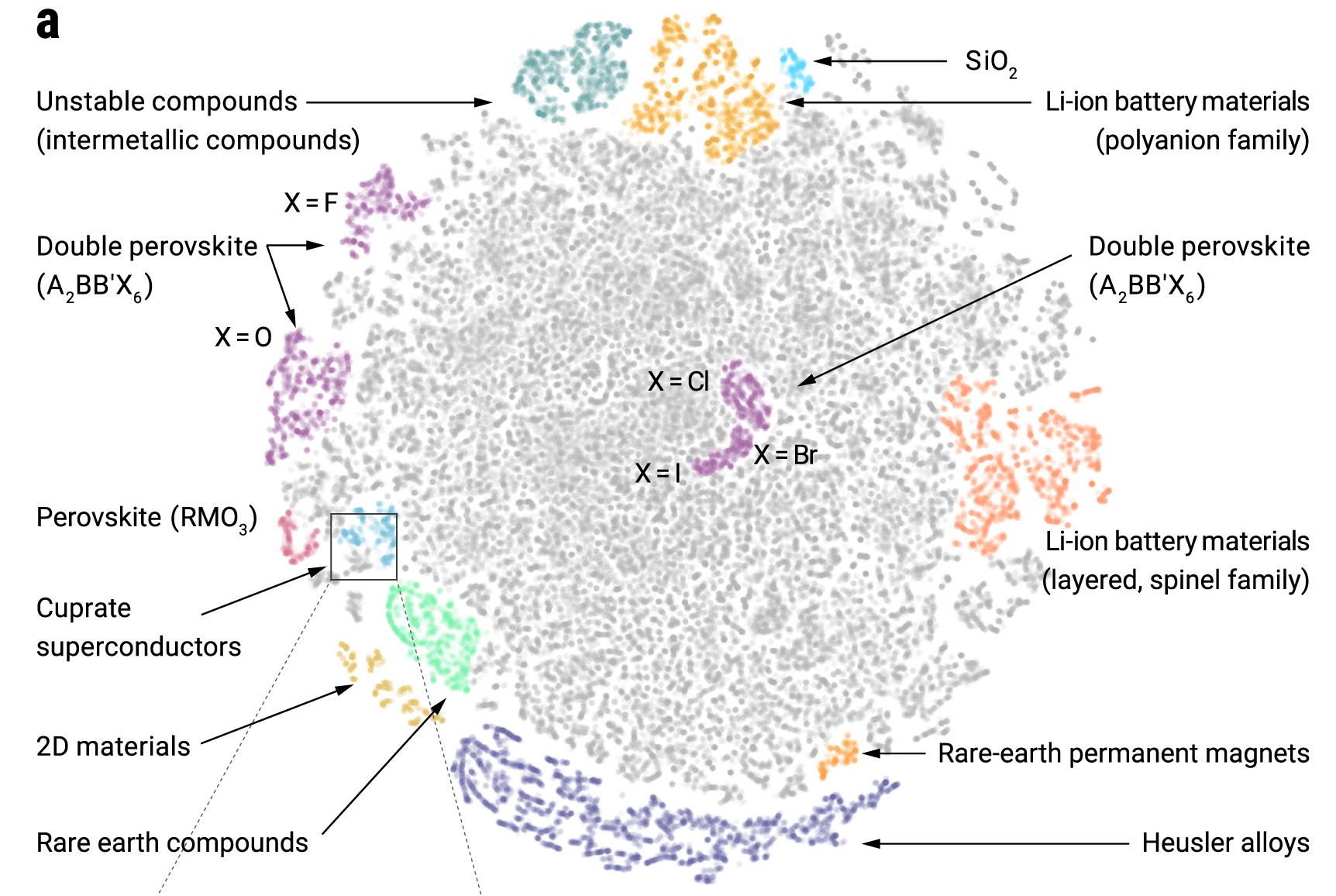}
    \caption{Figure adapted from Ref\cite{suzuki2022self}: A global map of the materials space of $122,543$ inorganic materials from the Materials Project plotted via a t-SNE visualisation of the embeddings. The map was annotated with cluster labels through manual inspection.}
    \label{fig:suzuki2022self}
\end{figure}

A different descriptors' featurization approach is employed by Suzuki et al.\cite{suzuki2022self}, since no physical knowledge is required throughout the process. Two parallel Neural Networks(NN) were used to embed $122,534$ inorganic crystals from the Materials Project (MP) into fixed-length feature vectors. These embeddings capture the local crystal structure, as well as periodicity with the X-ray diffraction (XRD) patterns computed using the Python materials analysis code ``pymatgen''. Unit cell atoms transformed into a graph feed a Crystal Graph Convolutional NN (CGCNN)\cite{xie2018crystal} while XRD patterns are used in a 1D Convolutional NN developed by \cite{park2017classification}. Each NN encoder generates $1024$-dimensional embedding vectors as output. t-SNE is applied to visualize the resultant embedding space [in Fig.\ref{fig:suzuki2022self}] (crystal structure together with XRD); target materials classes such as 2D materials, perovskites and cuprate semiconductors successfully cluster, supporting that this embedding space can capture structure-functionality relationship. The map is also used to see elemental, energy above hull (eV), bandgap (eV) and magnetization (T) distributions across the space, again highlighting regions of interest. Furthermore, local neighbourhood analysis is performed for the Hg-$1223$ superconductor, LiCoO$_2$ lithium-ion battery material, 2D ferromagnet Cr$_2$Ge$_2$Te$_6$, and Sm$_2$Co$_{17}$ permanent magnet using Euclidean distance. This analysis compares the embedding proposed in this study with those derived from Ewald sum and sine Coulomb matrices\cite{faber2015crystal}, showing superior performance of the proposed embedding. Furthermore, a supervised binary classifier for superconductors and thermoelectric materials built in the embedding space performed better than the one considered as a baseline \cite{xie2018crystal}, despite the reduced size of training sets, which is a transversal problem in the field. This work highlights the potential of neural networks' embedding capabilities, which in some cases outperform ``classical'' features.

\begin{figure}[t]
    \centering
    \includegraphics[width=1\linewidth]{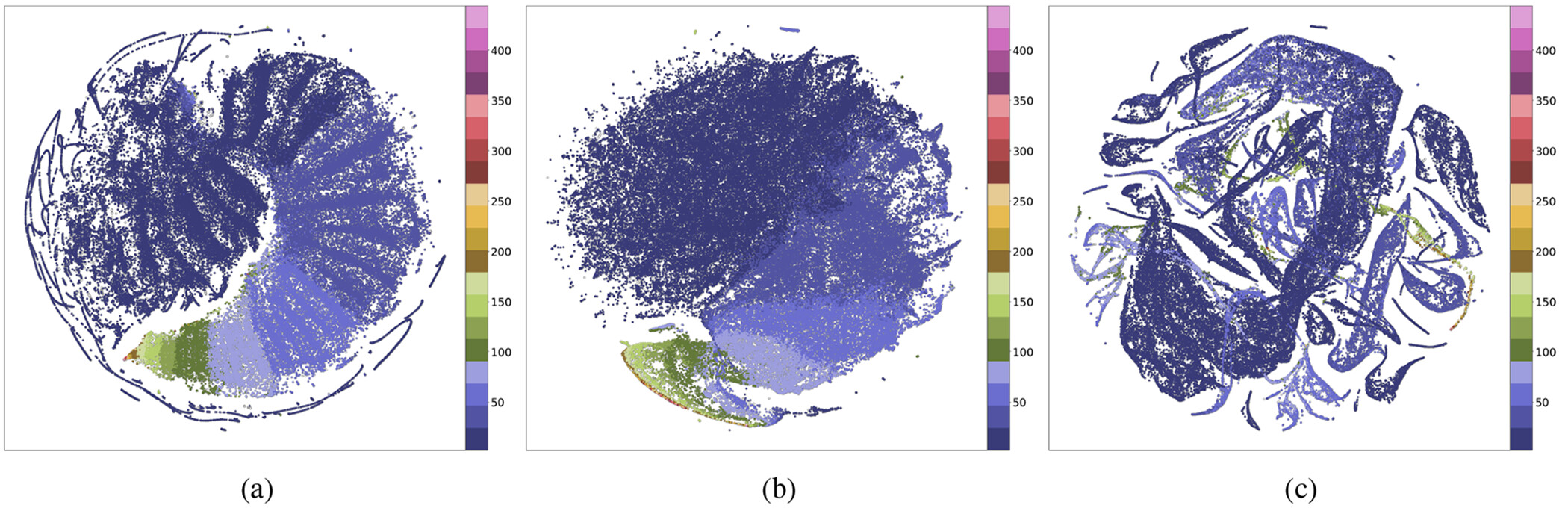}
    \caption{ "Distribution of materials in terms of atomic site numbers with different descriptors. The color corresponds to the property value map on the right of each graph. (a) Cartesian coordinates zero padding, (b) atomic pairwise distance, and (c) topology".\\
    \textit{Reprinted with permission from \cite{li2023global}. Copyright 2025 American Chemical Society}}
    \label{fig:li2023global}
\end{figure}

In another paper by Li et al.\cite{li2023global} the chosen dataset comprises $136,071$ crystal structures from that Material Project (MP) and seven different maps are created from seven different features extracted from compositional, structural, physical (XRD), topological, and latent NN's space descriptors; let us see them in detail. 
(i) {\em Atomic sites Cartesian coordinates}, and (ii) {\em fractional coordinates}, embedded with the zero-padding scheme, transform crystals with different numbers of atomic sites into fixed-length feature vectors. 
(iii) {\em Site pairwise distance matrix}: the distances within the complete data set are depicted in a frequency histogram divided into $100$ percentiles, allowing each material to be characterised by a $100$-dimensional vector, given its frequency distances within those intervals. The (iv){\em topological representation} of the crystal structure has been embedded into a $200$-dimensional feature via the atom-specific persistent homology\cite{jiang2021topological}(ASPH), which can capture both pairwise and many-body interactions. (v) {\em XRD spectra}, calculated using pymatgen\cite{ong2013python} from $0$ degrees to $89.98$ degrees with a step size of $0.1$ degrees, are then encoded into $900$-dimensional feature vectors. (vi){\em Elemental composition} is represented with one-hot encoding as a $2784$-dimensional vector, utilizing $32$ bits for each of the 87 elements included in the Material Project dataset under examination.
The (vii) {\em neural latent fingerprint} is extracted right before the last output layer of a DeeperGATGNN\cite{omee2022scalable}, trained for formation energy prediction using a masked MP materials ($36,837$ elements) training set for $500$ epochs with $20$ graph-convolution layers. 
The t-SNE has been utilized once more to visualize these maps, with various highlighted targets. In particular, factors such as atomic numbers, band gap (eV), density (g/cm\textsuperscript{3}), and formation energy per atom (eV/atom) for materials such as $ABC_{3}$ and $ABO_{3}$, chosen for their piezoelectric qualities, are depicted in these maps [Fig.\ref{fig:li2023global}].
It becomes evident that maps constructed from different features are more effective in highlighting specific material characteristics. Additionally to global mapping, it is possible to select a subset and dive into a local mapping, as has been done here for $ABC_3$ materials.  This work proposes a quality evaluation of clusters in mapping, examining the radius that encompasses the subset of target materials. For $ABC_3$ materials, the topology descriptor feature results as the most effective one.\\

\subsection*{Materials Networks}
\begin{figure}[t]
    \centering
    \includegraphics[width=1\linewidth]{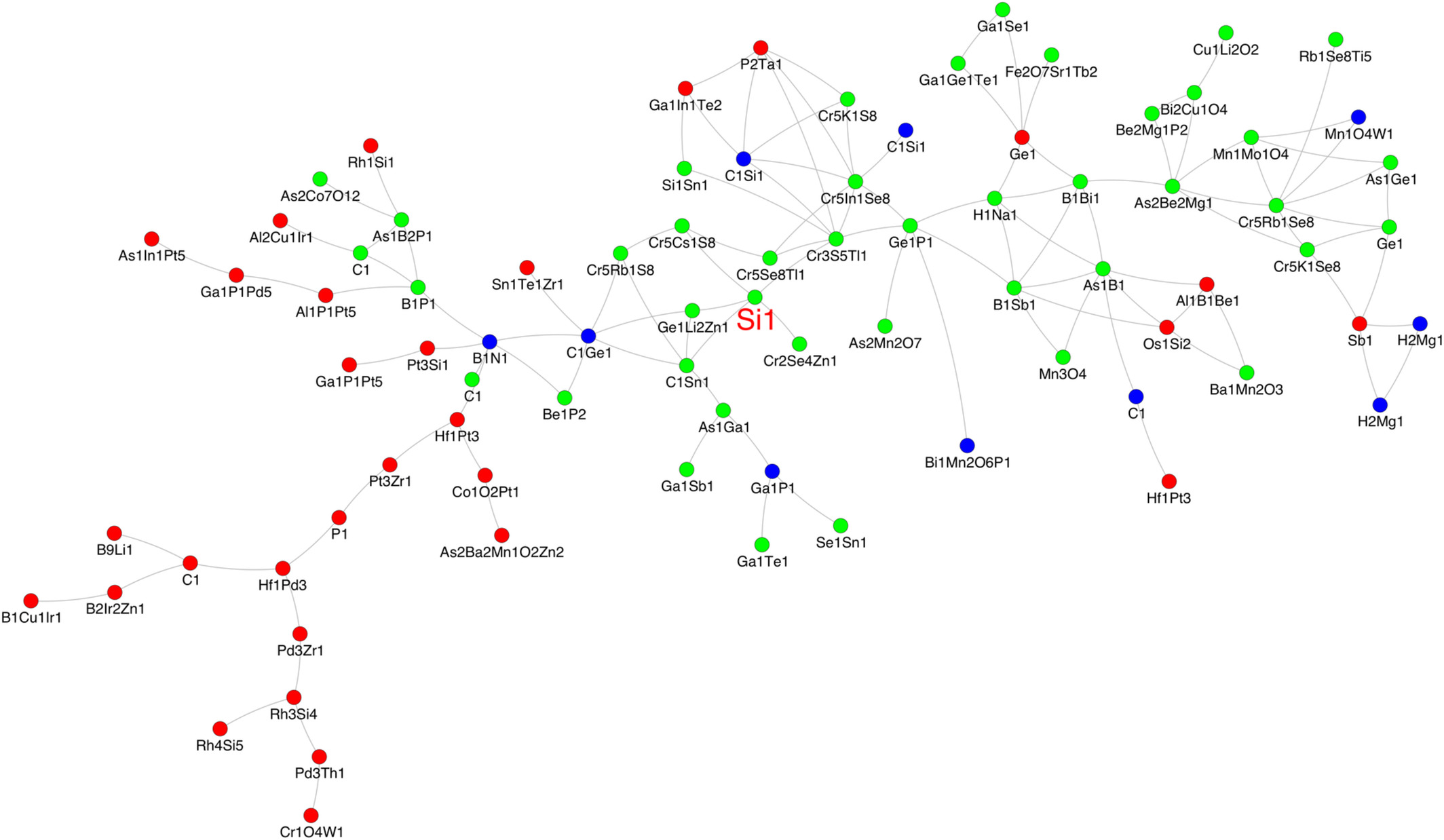}
    \caption{"Largest connected component of materials network that includes Si (ICSD 150530) containing 82 materials and 116 edges obtained for 0.82 cutoff of weighted Pearson coefficient with adjustment computed over negative energy region. The nodes are colored according to the band gap: red—metals (0 eV), green—semiconductors (0–1.5 eV), blue—insulators ($>$1.5eV)".\\
    \textit{Image reproduced with permission under License No. $5963090057385$. Copyright 2025, John Wiley and Sons - AIChE Jurnal}}
    \label{fig:veremyev2021networks}
\end{figure}

Isayev et al.\cite{isayev2015materials}
developed in 2014, a first layout of network-extracting framework. The authors worked on more than $20,000$ crystals from AFLOWLIB database, in which they created material cartograms and performed different analysis. They chose both $\Gamma$ point of the band structure and density of states as electronic structure descriptors and then developed a modified SiRMS descriptor to 'capture compositional, topological, and spatial (stereochemical) characteristics'\cite{isayev2015materials}. Once these descriptors have been encoded into vectors, they used Tanimoto similarity (or Jaccard index\cite{jaccard1901etude}) and a threshold $S=0.7$ to create a network. Two different cartograms/maps have been created from the two electronic structures descriptors. Specific classes of materials of interest (superconductors, topological insulators, pure, binary, ternary, etc.), as long as some of their properties (i.e. $T_c$) have been highlighted showing good/discrete clustering phenomena. A relevant result is the finding of a \textbf{scale-free} distribution of the network's edges. 
Considered later work, \cite{veremyev2021networks} is one more focused on network representation, exploring the effect of different similarity metrics and threshold values. Veremyev et al. analysed more than $27,000$ material's DOS functions, from the AFLOW repository, as unique descriptor. Following a comprehensive evaluation of various commonly employed similarity metrics and an assessment of their limitations, an adjusted Weighted Pearson correlation coefficient was formulated to enhance the physical coherence of the results. This methodology assigns greater significance to bands in closer proximity to the Fermi energy. Network analysis parameters (degree distribution, diameter, average distance, clustering coefficients, degree assortativity, maximum clique) are computed by varying the threshold, which determines the network structure, in particular, it fragments it into subgraphs. For example, the largest subgraph with $S = 0.82$ has $82$ components ($116$ edges), out of total nodes (more than $27,000$) [Fig\ref{fig:veremyev2021networks}]. Thus, \textbf{small-world} networks have been derived from the materials network in the subject.

\section*{Databases}


Generally, materials data come from measurements of synthesized materials or are the results of ab initio calculations and no strict division is presented amongst the two classes of data in the available databases. Structures are often collected by looking at scientific journal articles, if the structure appears in it, it should be possible to produce it or at least it might be stable.

We report a few of the largest databases available to map materials space and its subspaces. Here we shall not include organic or metal-organic structure databases (for which, by the way, the most used ones are the Cambridge Structural Database - CSD\cite{groom2016cambridge}, along with OMCSD and OMDB).

\subsubsection*{Crystal structure databases}
This section enumerates the databases primarily consisting of materials composition and crystal structures, from which the scientific community extracts data for subsequent analysis (clearly, a material is primarily defined by its composition; from now on it will not be specified, but compositional information is always available while talking of materials). 
\begin{itemize}
     \item \textbf{COD} - Crystallography Open Database: an open-access collection of more than $520,000$ crystal structures of \textit{organic, inorganic, metal-organic compounds and minerals, excluding biopolymers}. Crystal structures are collected from peer-reviewed papers; indeed, those are mainly obtained from experiments, although some may be predicted but well-validated. Due to its exclusive focus on structural data, the database maintains clarity and ease of access. Fully open access, no registration or access limitations are present.
     
    \item \textbf{ICSD} - Inorganic Crystal Structure Database\cite{zagorac2019recent}: curated database with more than $307,000$ inorganic crystal structures, subdivided as: more than $229,000$ experimental inorganic, more than $46,000$ metal-organic and more than $30,000$ theoretical (with low $E_{tot}$) and more than $27,000$ derived structures. Additionally, it offers simulation data of powder diffraction. Data are extracted from scientific journals by the editorial team, who subject them to quality checks. It can be accessed via desktop or web interface. The information in ICSD is updated biannually and to access it requires an annual subscription. Extensive documentation can be found on the ICSD website. 
    
    \item \textbf{MPDS} - Materials Platform for Data Science: based on PAULING FILE experimental inorganic database, again, extracted from peer-reviewed scientific journals. It contains more than $507,000$ structures. Although numerous properties have been computed in the MPDS, there is a lack of clear documentation on the exact number of materials for which these properties have been assessed. In contrast, comprehensive documentation is provided as tutorials for GUI and API interfaces, which also require different yearly subscription plans.

\end{itemize}

\subsubsection*{Ab-initio calculations databases}
From crystal structure to final calculated properties via ab-initio calculations; there are few platforms to do so, here are the most commonly utilized ones: 
    \begin{itemize}
        \item \textbf{Quantum ESPRESSO} - Quantum opEn-Source Package for Research in Electronic Structure, Simulation, and Optimization\cite{giannozzi2009quantum}: '\textit{is an integrated suite of Open-Source computer codes for electronic-structure calculations and materials modelling at the nanoscale. It is based on density-functional theory, plane waves, and pseudopotentials}'. Exhaustive documentation and tutorials are present on the website. 
        
        \item \textbf{ABINIT} ``\textit{(i)'s a package whose main program allows one to find the total energy, charge density and electronic structure of systems made of electrons and nuclei (molecules and periodic solids) within Density Functional Theory (DFT), using pseudopotentials (or PAW atomic data) and a planewave basis. ABINIT also optimize the geometry according to the DFT forces and stresses, or perform molecular dynamics simulations using these forces, or generate phonons, Born effective charges, and dielectric tensors, based on Density-Functional Perturbation Theory, and many more properties}''\cite{gonze2020bbinitproject}. This dataset is provided with clear documentation and tutorials and it is open-source. 
          
        \item \textbf{VASP} - Vienna Ab-initio Simulation Package\cite{kresse1996efficiency}. It computes electronic structure calculations and quantum-mechanical molecular dynamics from first principles. Additionally, optical, magnetic and phononic calculations could be performed. The platform is extensively documented and accessible due to its comprehensive tutorials. VASP is a commercial software with a paid license.

    \end{itemize}
    
High-throughput calculation managers are essential tools for automating, managing, and analyzing large-scale computational workflows in materials science. The most common high-throughput workflow managers include: AiiDa, FireWorks, pymatgen, Atomate and AFLOW. These tools are essential for performing ab-initio calculations on extensive structure databases, producing the properties we seek. 

\subsubsection*{Calculated properties databases}
Materials informatics derives data predominantly from these databases. Consequently, the most readily available databases tend to be the most frequently utilized. Furthermore, data retrieval has been simplified by the presence of well-designed web interfaces and APIs.
    \begin{itemize}
        \item \textbf{Material Project} \cite{jain2013commentary}: This is undoubtedly the most frequently utilized resource. Built with VASP and pymatgen high-throughput workflow, it includes many calculated properties. Through the web API, it becomes apparent that as we request more properties at once, the number of available materials diminishes. Starting from the entire inorganic structures database $\sim170k$, it reduces to 50k+ with BS, DOS and magnetic properties, contracts further to 11k+ when including elasticity properties, and drops to $\sim2000$ materials when dielectric information is desired (easy to verify with online API; MP contains many other information). The platform features a well-organized interface along with comprehensive documentation and tutorials. Completely open access with registration. 
        
        \item \textbf{AFLOWLIB} \cite{curtarolo2012aflowlib}: is an open-access database that stores results generated by the AFLOW framework. The web API provides easy access to the data, primarily featuring material structures, their formation enthalpies and band structures (360k+). In addition, it includes information on the thermal and elastic properties($<10k$). Extensive documentation and online lectures. 
        
        \item \textbf{OQMD} - Open Quantum Materials Database: is a database of DFT-calculated thermodynamic and structural properties of 1,226k+ materials. The stability, band gap, and formation energy are computed for each configuration. The website offers visualization of phase diagrams and structures. The APIs and the DFT parameters used are well documented.
        
        \item \textbf{NoMaD} - Novel Materials Discovery\cite{scheidgen2023nomad}:in addition to the extensive dataset available, the platform provides the capability to upload and analyze user-specific data. The VASP package is primarily utilized to perform calculations within NOMAD. The rapid advancement of the platform indicates potential for further expansion in the near future.

        \item \textbf{Materials Cloud}: based primarily on Aiida and Quantum Espresso, is an open platform to computational material science. The flagship databases, MC3D\cite{huber2022materialscloud} and MC2D\cite{mounet2020twodimensional} contain more than 34,000 3D crystals and 3,000 2D crystals, respectively; many more specific databases are available. API and documentation, along video workshop, are provided. 
\end{itemize}

The fundamental effort of unifying access to multiple materials databases is being made by Open Databases Integration for Materials Design (OPTIMADE\cite{evans2024developments}) consortium, which '\textit{aims to make materials databases interoperable by developing a specification for a common REST API}'. A total of 25 providers with 29 databases, comprising the above-cited ones, can be accessed via OPTIMADE. Once set up various APIs keys and accounts, the workflow seems quite effortless with respect to interact with each single database API. OPTIMADE embodies the desire and necessity of unified, coherent and easy-to-access databases.

\section*{Conclusions and Outlook}
From the work presented so far, the crucial role that emergent properties of chemical space play in understanding the structure of the space itself and its subsets appears. To interact with such emergent phenomena, we need more data, an accurate sampling of the space but above all a deep understanding of the nature of the descriptors and their non-linearity. Until now, unfortunately, the only databases that contain a sufficient amount of data (more than $10^6$ materials) are those based on crystal structure, which is why the community has mainly focused on deriving representations from this descriptor.

Although the relationship structure-properties is strong, probably it will not be sufficient to describe the chemical space complexity and to represent in its entirety the structure of chemical space. For this reason some works are using, as an example, electronic structure (or DOS) as descriptors, which is a first step in mapping materials by their properties. Mapping materials by properties could capture those non-linearities that we are seeking for technological developments. 

At the same time, it is also fair to conclude that the field of exploring chemical space with machine learning and network techniques is growing at an increasing pace.
The size of available databases will certainly advance in the number of their elements and in the quality of information stored. In particular, completeness, coherence, unification and many other aspects are and will be even more central in the near future, since ML and material informatics need as much good data as possible to increase performance in classification and prediction.
A promising step forward in achieving this will probably be related to the introduction of A-labs (Automatic laboratories), which aim to automate the materials synthesis and their
characterisation.
The final contact point between the measured and the synthetic databases will be a reciprocal validation, that will benefit both theoretical models and synthesis techniques. 
 
We are observing and testifying the interest and the movement of the community to improve databases by size, coherence, quality and variety of calculated properties. Since this will be an inevitable direction, we seek for those works defining techniques and tools that will be more valuable in the future than now.
Open problems to work on: databases ab initio calculations, minimum set of descriptors with maximal information without redundancy, featurization with less information loss as possible in compact numerical low-dim. vectors, which dimensionality reduction is best for which set of features, internal (similarity) metrics exploration, set of targets or parameters to evaluate the mapping process.

\end{document}